\documentclass[
aps,
prc,
twocolumn,
showpacs,
superscriptaddress,
nofootinbib,
floatfix]
{revtex4}
\usepackage{graphicx,
longtable}

\def\be{\begin{equation}}
\def\ee{\end{equation}}
\newcommand{\dg}{^{\dagger}}
\def\bee{\begin{eqnarray}}
\def\eee{\end{eqnarray}}
\newcommand{\ket}[1]{\left|#1\right\rangle}
\newcommand{\bra}[1]{\left\langle #1\right|}

\newcommand{\lf}{\left(}
\newcommand{\rh}{\right)}

\begin{document}

\title{Two-neutrino double-beta decay Fermi transition and two-nucleon interaction}
%
%
\author{Du\v{s}an \v{S}tef\'anik}
\affiliation{ Comenius University, Mlynsk\'a dolina F1, SK--842 48, Slovakia}
\author{Fedor \v{S}imkovic}
\affiliation{ Comenius University, Mlynsk\'a dolina F1, SK--842 48, Slovakia}
\affiliation{ BLTP, JINR, 141980 Dubna, Moscow region, Russia}
\affiliation{ IEAP CTU, 128--00 Prague, Czech Republic}
\author{Kazuo Muto}
\affiliation{Department of Physics, Tokyo Institute of Technology, Tokyo 152-8551, Japan }				
\author{Amand Faessler}
\affiliation{ Institute of Theoretical Physics, University of Tuebingen,72076 Tuebingen, Germany}

\begin{abstract}
An exactly solvable model for a description of the two-neutrino double beta decay transition of the Fermi 
type is considered. By using perturbation theory an explicit dependence of the
two-neutrino double beta decay matrix element on the like-nucleon pairing, particle-particle 
and particle-hole proton-neutron interactions by assuming a weak violation of isospin symmetry 
of Hamiltonian expressed with generators of the SO(5) group.
It is found that there is a dominance of double beta decay transition through a single state 
of the intermediate nucleus. Then, an energy weighted sum rule connecting $\Delta$Z=2 nuclei 
is presented and discussed. It is suggested that this sum rule  can be exploited to study 
the residual interactions of the nuclear Hamiltonian.
\end{abstract}
\medskip

\pacs{
21.60.Fw, 21.60.Jz, 23.40.Hc}
\maketitle

%
\section{Introduction}
%

The two-neutrino double-beta decay ($2\nu\beta\beta$-decay),
which involves the emission of two electrons and two antineutrinos
\cite{doi83,haxton,suci98,fs98,ves12} 
\begin{equation}
(A,Z) \rightarrow (A,Z+2) + 2 e^- + 2 {\tilde{\nu}}_e,
\end{equation}
has attracted the attention of both experimentalists and theoreticians
for a long period and remains of major importance for nuclear physics. 

It is a  second order process in the weak interaction allowed in
the standard model. The $2\nu\beta\beta$-decay can be observed, because 
due to the pairing force even-even nuclei with an even number of 
protons and neutrons are more stable than the odd-odd nuclei
with broken pairs \cite{doi83,haxton}. Thus, the  single $\beta$-decay  transition  
from the (A,Z) nucleus to neighboring odd-odd nucleus is 
energetically forbidden. 

Till now, the $2\nu\beta\beta$-decay has been detected for 11
different nuclei for transition to the ground state and in two 
cases also to transition to $0^+$ excited state of the daughter
nucleus \cite{recbb}. This rare process is one of the major sources of 
background in running and planned experiments looking for a 
signal of the more fundamental neutrinoless double-beta decay, which 
occurs if the neutrino is a massive Majorana particle.  

The inverse half-life of the $2\nu\beta\beta$-decay is free 
of unknown parameters of particle physics and can
be factorized to a good approximation as \cite{doi83,haxton}
\begin{equation}
\left( T^{2\nu}_{1/2}\right)^{-1} = G^{2\nu} g^4_A
\left| M^{2\nu}_{GT} - \left(\frac{g_V}{g_A}\right)^2 M^{2\nu}_F\right|^2,
\end{equation}
where $G^{2\nu}$ is the lepton phase-space factor, $g_A$ ($g_V$) is the 
axial-vector (vector) coupling constant. The $2\nu\beta\beta$-decay 
is governed by the double Gamow-Teller (GT) and double Fermi (F) matrix 
elements, which are given by \cite{doi83,haxton,suci98,fs98} 
\begin{eqnarray}
M^{2\nu}_{F, GT} &=& 
\sum_n \frac{\langle f \parallel {\cal O}_{F,GT} \parallel 1^+_n \rangle
\langle 1^+_n \parallel {\cal O}_{F,GT} \parallel i \rangle}
{E_n - (E_i + E_f)/2}
\end{eqnarray}
with
\begin{eqnarray}
{\cal O}_F = \sum_{k=1}^A \tau^+_k,~~~~~  
{\cal O}_{GT} = \sum_{k=1}^A \tau^+_k \boldmath{\sigma_k}.
\end{eqnarray}
where $|i>$ ($|f>$) are $0^+$ ground states of the initial (final)
even-even nuclei with energy $E_i$ ($E_f$), 
and $|1^+_n>$ ($|0^+_n>$) are the $1^+$ ($0^+$) states in the 
intermediate odd-odd nucleus  with energies $E_n$. 

Many attempts have been made in the literature to calculate the 
$2\nu\beta\beta$-decay nuclear matrix elements (NMEs) for nuclei 
of experimental interest \cite{doi83,haxton,suci98,fs98,domin,rath07,sarr}. 
Recent results obtained within 
the nuclear shell model are in a good agreement with the measured 
$2\nu\beta\beta$-decay half-lives \cite{poves12}. But, 
it is achieved  by a consideration of significant 
quenching by a factor q=0.4-0.7 of the Gamow-Teller operator, 
which is obtained by a  normalization
of the total theoretical $\beta^-$ strength in the experimental energy window 
to the  measured one. 

The  quasiparticle random phase approximation (QRPA) has been found to be
successful in revealing the suppression mechanism for the $2\nu\beta\beta$-decay
NMEs \cite{vogel86,civitarese87,muto89}. However, the predictive power of the 
QRPA is questionable because of extreme sensitivity of calculated $2\nu\beta\beta$-decay
matrix elements in the physically acceptable region on the particle-particle
strength of nuclear Hamiltonian. In \cite{muto89} it was shown that 
if this strength is determined from a QRPA calculation of single $\beta^+$ decays
a reasonable agreement with the measured $2\nu\beta$-decay is achieved.

The quenching behavior of the $2\nu\beta\beta$-decay matrix elements is 
a puzzle and has attracted the attention of many theoreticians. Recently, 
it was shown that $M^{2\nu}_F$ depends strongly on the isovector part 
of the particle-particle neutron-proton interaction unlike $M^{2\nu}_{GT}$,
which  depends strongly on its isoscalar part \cite{newpar13}.  
The underlying symmetries responsible for these suppressions are assumed 
to be isospin SU(2) and spin-isospin SU(4) symmetries in the cases of double Fermi and 
double Gamow-Teller NMEs, respectively \cite{desplan90}. 
 
The goal of this paper is to discuss the suppression mechanism of the double Fermi
matrix element close to the point of restoration of isospin symmetry of the nuclear 
Hamiltonian in the context of residual nucleon-nucleon interaction. For the sake of 
simplicity we consider a schematic Hamiltonian, describing the gross properties 
of the beta-decay processes in the simplest case of monopole Fermi transitions 
within the SO(5) model \cite{kuzmin,muto92,civi94,hplb,hirsch97,krmpotic98}. 
In order to find explicit dependence 
of $M^{2\nu}_F$ on different parts of the nuclear Hamiltonian the perturbation theory 
is exploited. We note that the SO(5) model remains
a tool for understanding of different nuclear physics phenomena 
even nowadays \cite{engel04,anatomy,engel12}.

%
\section{Schematic Hamiltonian within the  SO(5) model}
%

In the model, protons and neutrons occupy only a single j-shell. The Hamiltonian includes
a single-particle term, proton-proton and neutron-neutron pairing, and a  
charge-dependent two-body interaction with both particle-hole and particle-particle
channels as follows:
\begin{eqnarray}
H &=& e_pN_p+e_nN_n-G_p S_p\dg S_p-G_n S_n\dg S_n \nonumber \\
&+&2\chi\beta^-\beta^+-2\kappa P^- P^+, \label{ham}
\end{eqnarray}
where
\begin{eqnarray}
&&N_{i}=\sum_m a\dg_{m,t_i} a^{\phantom{\dagger}}_{m,t_i}, 
\quad \beta^{-}=\sum_m a\dg_{m,-\frac{1}{2}} a^{\phantom{\dagger}}_{m,\frac{1}{2}},  \nonumber \\
&& S_{i}\dg = \frac{1}{2} \sum_m a\dg_{m,t_i} {\tilde a}\dg_{m,t_i}, \quad
P^{-}=\sum_m a\dg_{m,-\frac{1}{2}} {\tilde a}^{\dagger}_{m,\frac{1}{2}}, \nonumber \\
\end{eqnarray}
with i=p, n and $t_{n,p}=\pm 1/2$.
$a\dg_{m t}$ ($a_{m t}$) is creation (annihilation) operator of single
particle state $|jm,t>$ for protons and neutrons ($t = t_p , t_n$) and
${\tilde a}\dg_{m t} = (-1)^{j-m} a\dg_{-m t}$.

We rewrite  Hamiltonian (\ref{ham}) with help of operators
\begin{eqnarray}
&& A\dg\lf T_z \rh= \frac{1}{\sqrt{2}}\left[a\dg\otimes a\dg \right]^1_{T_z}, \nonumber \\
&& N = N_p+N_n, \quad  T_z=\frac{N_n-N_p}{2}, \nonumber\\
&& T^{-}=-\sqrt{2\Omega}\sum_m a\dg_{m,-\frac{1}{2}} a^{\phantom{\dagger}}_{m,\frac{1}{2}}.
\label{operators}
\end{eqnarray}
Here, $A\dg\lf T_z \rh$ is the nucleon pair creation operator with angular momentum $J=0$,
isospin $T=1$ and its projection on z-axis $T_z$ ($T_z = 0,\pm 1$). $N$, $T_z$ and $T^{-}$  
are the particle-number operator, the isospin projection and the isospin lowering operators,
respectively. It holds the identity $T^2=\lf T^{-}T^{+}+T^{+}T^{-}\rh/2 +T_z^2$.
$\Omega=j+1/2$ denotes the  semi-degeneracy of the considered single level.
The operators (\ref{operators}) with their Hermitian conjugates represent 
ten generators of the SO(5) group \cite{parikh}. We assume, the system is in seniority s=0. 
Then,  $[A^{\dg}\tilde{A}]_0^{0}$ expressed with  the SO(5) Casimir operator \cite{parikh} 
is given by
\begin{eqnarray}
[A^{\dg}\tilde{A}]_0^{0}=\frac{1}{2\sqrt{3}\Omega}\left[ \lf 2\Omega+3-N/2 \rh N/2-T(T+1) \right]. 
\end{eqnarray}

For the Hamiltonian (\ref{ham}) we get 
\begin{widetext}
\begin{eqnarray}
H&=&\left[e_n + e_p- \frac{1}{3}\left(3+2\Omega-
\frac{N}{2}\right)\left(\frac{G_p+G_n}{2}+2\kappa\right)\right]\frac{N}{2}
+\left[e_n - e_p-2\chi(T_z+1)\right]T_z  \nonumber \\
&&+\left[2\chi+\frac{1}{3}\left(\frac{G_p+G_n}{2}+2\kappa\right)\right]T(T+1)  \nonumber \\
&&+\frac{\Omega}{\sqrt{2}}\left(\frac{G_p-G_n}{2}\right)[A^{\dg}\tilde{A}]^1_{0}
+\sqrt{\frac{2}{3}}\Omega\left(4\kappa-\frac{G_p+G_n}{2}\right)[A^{\dg}\tilde{A}]^2_{0}. 
\label{hamupraveny}
\end{eqnarray}
\end{widetext}
As a consequence of the presence of the isovector and isoquadrupole terms
in Hamiltonian (\ref{hamupraveny}) the isospin is not  conserved in general.
It is due to differences between proton and neutron pairing strengths and an arbitrary 
strength of the proton-neutron isovector pairing component. However, particle number and 
isospin projection remains as good quantum numbers. 

The k$^{th}$ eigenstates of the Hamiltonian (\ref{hamupraveny})  with quantum numbers
N and T$_{z}$ can be expressed in terms of a basis labeled by a chain of 
irreducible representations of the SO(5) group 
(see Appendix \ref{ap}), namely 
\begin{equation}
|k; N T_z\rangle = \sum_T c^{(k)}_{N T T_z} |N T T_z\rangle. 
\end{equation}
A diagonalization of H requires calculation of matrix elements 
$\bra{N,T,T_Z}H\ket{N,T,T_Z}$ and $\bra{N,T\pm2,T_Z}H\ket{N,T,T_Z}$. 
The corresponding reduced matrix elements are given Appendix \ref{ap}).
For $G_p=G_n$ and $(G_p+G_n)/2=4\kappa$ the Hamiltonian (\ref{hamupraveny})
is diagonal in the basis of states $|N,T, T_z\rangle$.

%
\section{Double Fermi matrix element within perturbation theory}
%

We shall assume a small violation of the isospin symmetry due to isotensor term
of nuclear Hamiltonian (\ref{hamupraveny}). For the numerical example we consider 
a large value of j to simulate the realistic situation corresponding to
medium- and heavy-mass nuclei. The parameters chosen are given by
\be
\begin{array}{ccc}
 \Omega=10,          &   N=20,               & 1\leq T_z \leq 5,   \\
 e_p=0.3~\text{MeV}, &   e_n=0.1~\text{MeV}, & G = 0.165~\text{MeV}, \\
 G_p = G_n = G,      & \chi=0.044~\text{MeV}, & 0.7 \le 4 \kappa/G \le 1.3.
\end{array} \label{parameters}
\end{equation}

\begin{figure}[!t]
\begin{center}
\includegraphics[width=85mm]{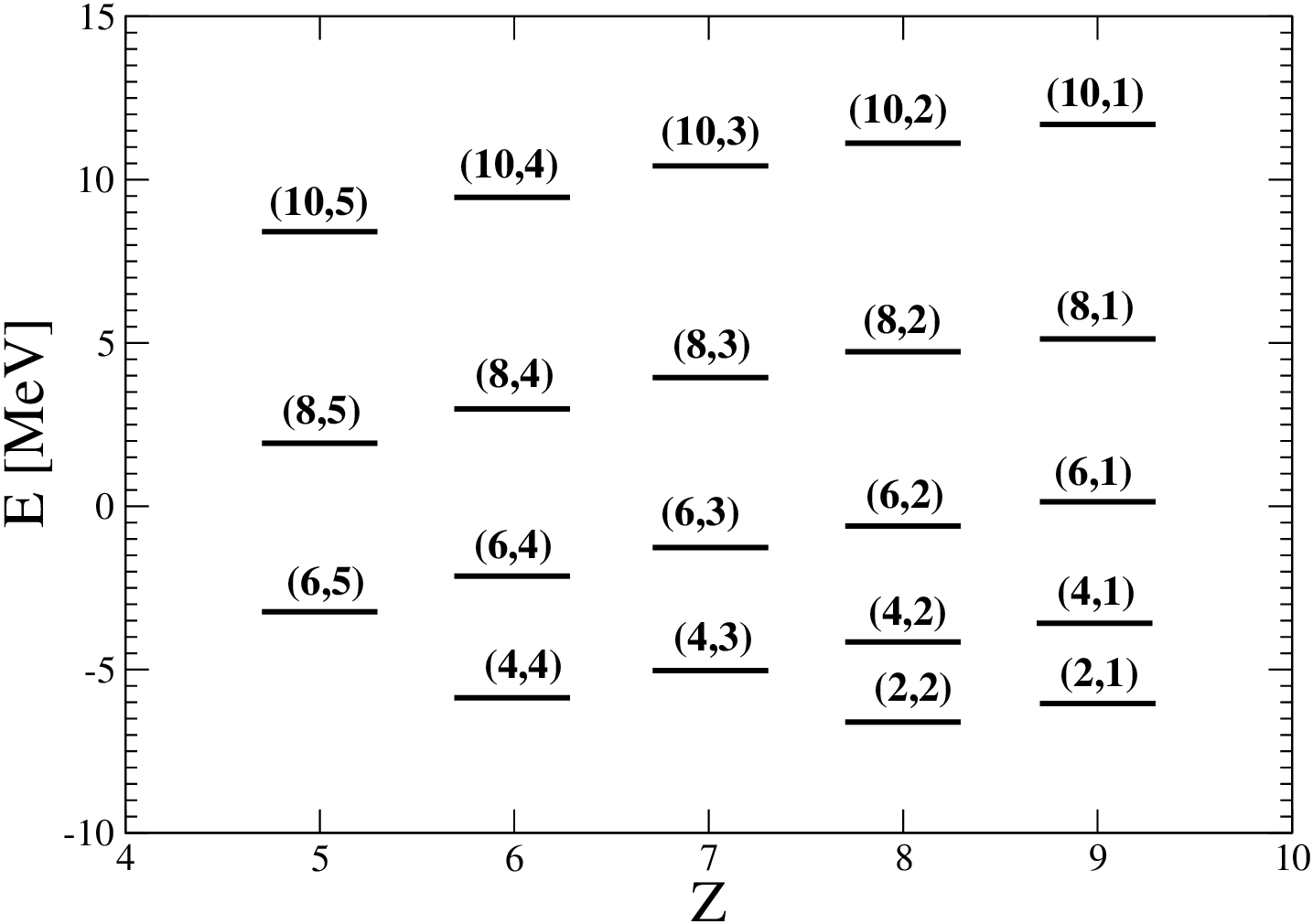}
\caption{The energy of the $0^{+}$ states of different isotopes are shown for
j=19/2 (and the set of parameters (\ref{parameters}) with $4 \kappa/G =1$)  
in MeV vs. Z plot. States are labeled by $\lf T, T_z \rh$.
\label{elevs}}
\end{center}
\end{figure}

For $4 \kappa/G =1$ the isospin symmetry is restored.  In Fig. \ref{elevs}
we present $0^+$ states with energy $E_{TT_z}$ of different isotopes. This level 
scheme illustrates the situation for the $2\nu\beta\beta$-decay of $^{48}$Ca.
The isospin  is known to be, to a very good approximation, a valid quantum 
number in nuclei. The ground states of $^{48}$Ca and $^{48}$Ti
can be identified with T=4 $T_z=4$ and T=2  $T_z=2$, respectively, i.e.
they are assigned into different isospin multiplets. As the total isospin projection
lowering operator $T^-$  is not changing the isospin the double Fermi 
matrix element $M^{2\nu}_F$ is non-zero only to  the extent that the Coulomb
interaction mixes the high-lying T=4 $T_z$=2 analog of the $^{48}$Ca ground
state into the T=2  $T_z=2$ ground state of $^{48}$Ti. 

We shall study double Fermi matrix element in the perturbation theory within the
discussed model close to a point of a restoration of the isospin symmetry
($4 \kappa/G =1$). The isoscalar and isotensor terms of the Hamiltonian 
(\ref{hamupraveny}) represent the unperturbated and perturbated terms, respectively.
We denote perturbated states and their energies with a superscript
prime symbol ($|T' T_z\rangle$, $E'_{TT_z}$) unlike the states with a 
definite isospin ($|T T_z\rangle$, $E_{TT_z}$). Up to the second order
of parameter $(4\kappa-G)$ we find
\begin{eqnarray}
E'_{44}&=&14 e_n+6e_p-\frac{110}{3}\lf G+2\kappa \rh \nonumber \\ \nonumber
&& -\sqrt{\frac{2}{3}}\Omega\lf G-4\kappa\rh\bra{44}[A^{\dg}\tilde{A}]_0^2\ket{44} \\
&& -\frac{2}{3}\Omega^2\lf G-4\kappa\rh^2\frac{\bra{64}[A^{\dg}
\tilde{A}]_0^2\ket{44}^2}{44\chi+22/3\lf G+2\kappa \rh},
\label{ig}\\
E'_{43}&=&13 e_n+7e_p+16\chi-\frac{110}{3}\lf G+2\kappa \rh \nonumber \\ \nonumber
&& -\sqrt{\frac{2}{3}}\Omega\lf G-4\kappa\rh\bra{43}[A^{\dg}\tilde{A}]_0^2\ket{43} \\
&& -\frac{2}{3}\Omega^2\lf G-4\kappa\rh^2\frac{\bra{63}[A^{\dg}
\tilde{A}]_0^2\ket{43}^2}{44\chi+22/3\lf G+2\kappa \rh},
\label{ng}\\
E'_{22}&=& 12 e_n+8e_p-\frac{124}{3}\lf G+2\kappa \rh \nonumber \\ \nonumber
&& -\sqrt{\frac{2}{3}}\Omega\lf G-4\kappa\rh\bra{22}[A^{\dg}\tilde{A}]_0^2\ket{22} \\
&& -\frac{2}{3}\Omega^2\lf G-4\kappa\rh^2\frac{\bra{42}[A^{\dg}
\tilde{A}]_0^2\ket{22}^2}{28\chi+14/3\lf G+2\kappa \rh}, 
\label{fg}\\
E'_{42}&=&12 e_n+8e_p+28\chi-\frac{110}{3}\lf G+2\kappa \rh \nonumber \\ \nonumber
&& -\sqrt{\frac{2}{3}}\Omega\lf G-4\kappa\rh\bra{42}[A^{\dg}\tilde{A}]_0^2\ket{42} \\ \nonumber
&& +\frac{2}{3}\Omega^2\lf G-4\kappa\rh^2
\frac{\bra{42}[A^{\dg}\tilde{A}]_0^2\ket{22}^2}{28\chi+14/3\lf G+2\kappa \rh} \\
&& -\frac{2}{3}\Omega^2\lf G-4\kappa\rh^2
\frac{\bra{62}[A^{\dg}\tilde{A}]_0^2\ket{42}^2}{ 44\chi+22/3\lf G+2\kappa \rh}. 
\label{fe}
\end{eqnarray}
The particular matrix elements of SO(5) operators connecting states with a definite 
isospin and its projection are presented in Appendix \ref{ap}).

For transition $|4'4> \rightarrow |2'2>$ the double Fermi matrix element 
can be written as
\begin{equation}\label{fermione}
M^{2\nu}_F = \sum_{T=4,6,8,10}^{10}\frac{\bra{2' 2}T^-\ket{T' 3}\bra{T' 3}T^-\ket{4' 4}}
{E'_{T 3}-\lf E'_{4 4}-E'_{2 2}\rh/2}.
\label{medfermi}
\end{equation} 
It contains a sum over the states of the intermediate nucleus $|T' 3\rangle$. However,
up to second order of perturbation theory there is only a single 
contribution through the intermediate state $|4' 3\rangle$. Thus, we have
\begin{equation}
M^{2\nu}_F \simeq \frac{\bra{2' 2}T^-\ket{4' 3}\bra{4' 3}T^-\ket{4' 4}}
{E'_{3 3}-\lf E'_{4 4}-E'_{2 2}\rh/2}.
\label{appmedf}
\end{equation} 
The involved $\beta$-transition amplitudes are given by
\begin{widetext}
\begin{eqnarray}
\bra{4' 3}T^-\ket{4' 4}
&=&\bra{43}T^{-}\ket{44} \lf 1-\frac{1}{3}
\frac{\Omega^2\lf 4\kappa-G \rh^2}{\lf 44\chi+22/3\lf G+2\kappa \rh\rh^2} 
\left[\left|\bra{44}[A^{\dg}\tilde{A}]_0^2\ket{64}\right|^2
+\left|\bra{43}[A^{\dg}\tilde{A}]_0^2\ket{63}\right|^2\right]\rh 
\label{betaminus} \\ \nonumber
&& + \bra{63}T^{-}\ket{64}\frac{2}{3}
\frac{\Omega^2\lf 4\kappa-G \rh^2}{\lf 44\chi+22/3\lf G+2\kappa \rh\rh^2}
\bra{64}[A^{\dg}\tilde{A}]_0^2\ket{44}\bra{63}[A^{\dg}\tilde{A}]_0^2\ket{43}
\end{eqnarray}
and
\begin{eqnarray}
\bra{2' 2}T^-\ket{4' 3}
&=&\bra{42}T^{-}\ket{43}\left[\sqrt{\frac{2}{3}}\Omega\lf G-4\kappa 
\rh\frac{\bra{42}[A^{\dg}\tilde{A}]_0^2\ket{22}}
{\lf 28\chi+14/3\lf G+2\kappa \rh\rh}\right. \label{betaplus} \\ \nonumber
&& +\left.\frac{2}{3}\frac{\Omega^2\lf G-4\kappa\rh^2}{\lf 28\chi+14/3\lf G+2\kappa 
\rh\rh^2}\lf  \bra{42}[A^{\dg}\tilde{A}]_0^2\ket{42}\bra{42}[A^{\dg}\tilde{A}]_0^2\ket{22}
-\bra{22}[A^{\dg}\tilde{A}]_0^2\ket{22}\bra{42}[A^{\dg}\tilde{A}]_0^2\ket{22}\rh \right].
\end{eqnarray}
\end{widetext}
If isospin symmetry is restored ($4\kappa = G$) we end up with
$\bra{2' 2}T^-\ket{4' 3}=\bra{22}T^-\ket{43}=0$. For the
energy denominator in (\ref{appmedf}) with help of Eqs. (\ref{ig}), 
(\ref{ng}) and (\ref{fg}) we get
\begin{widetext}
\begin{eqnarray}\label{enemen}
&& E'_{43} - \lf E'_{44} - E'_{22}\rh/2 = 16\chi +\frac{7}{3}\lf G+2\kappa\rh\\ \nonumber
&&+\sqrt{\frac{1}{6}}\Omega\lf4\kappa-G\rh\left[2\bra{43}[A^{\dg}
\tilde{A}]_0^2\ket{43}-\bra{44}[A^{\dg}\tilde{A}]_0^2\ket{44}-\bra{22}[A^{\dg}
\tilde{A}]_0^2\ket{22}\right] \\ \nonumber
&&+\frac{1}{3}\Omega^2\lf 4\kappa-G\rh^2
\left[\frac{\bra{64}[A^{\dg}\tilde{A}]_0^2\ket{44}^2}{44\chi+22/3\lf G+2\kappa \rh}
+\frac{\bra{42}[A^{\dg}\tilde{A}]_0^2\ket{22}^2}{28\chi+14/3\lf G+2\kappa \rh}
-2\frac{\bra{63}[A^{\dg}\tilde{A}]_0^2\ket{43}^2}{44\chi+22/3\lf G+2\kappa \rh}\right] 
\end{eqnarray}
\end{widetext}
We note that the energy denominator $E'_{43}-\lf E'_{44}-E'_{22}\rh /2$ as well 
as the whole double Fermi matrix element $M^{2\nu}_F$ does not depend explicitly 
on mean field  parameters $e_p$ and $e_n$. 

\begin{figure}[!t]
\begin{center}
\includegraphics[width=85mm]{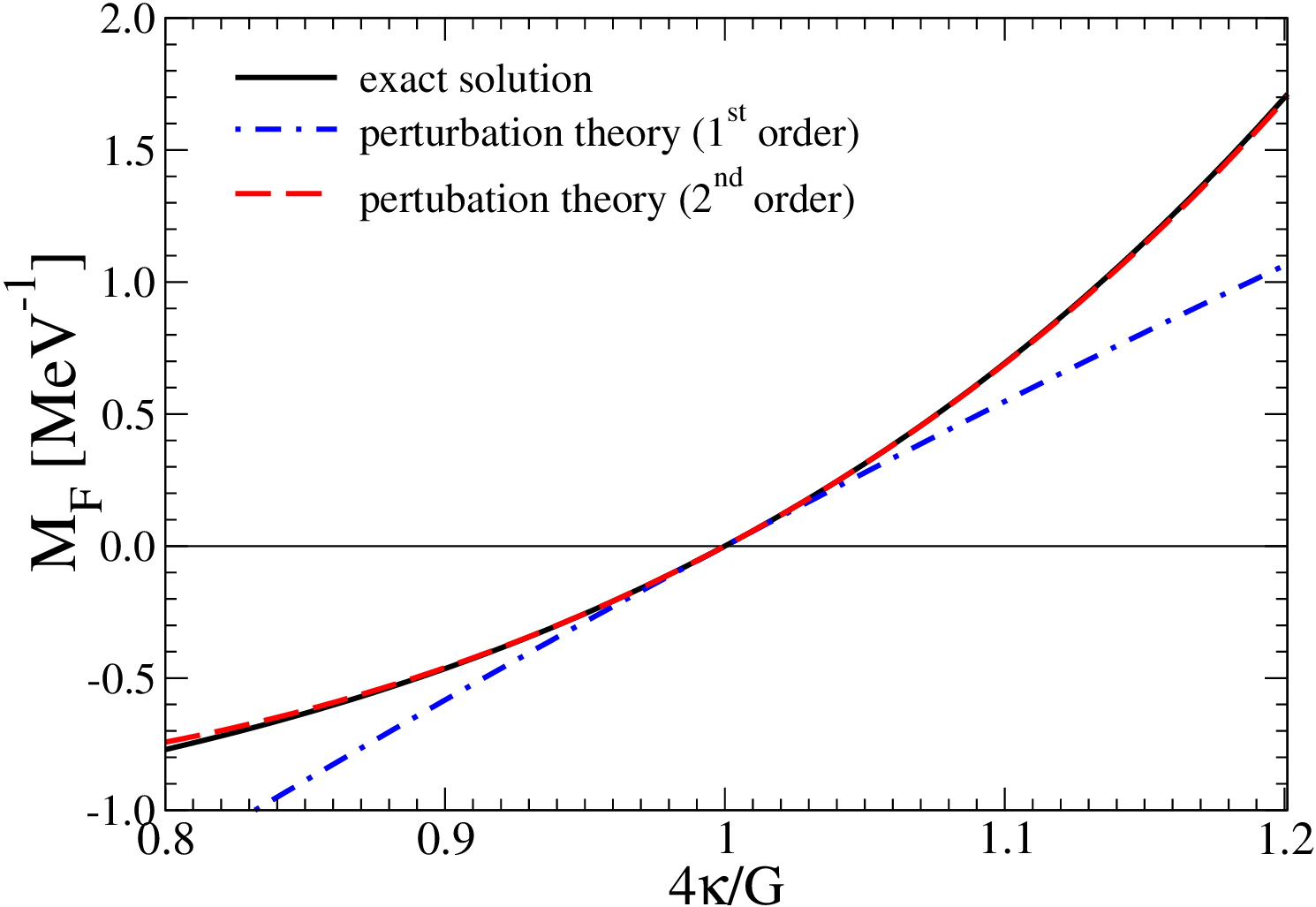}
\caption{(Color online) Matrix element $M^{2\nu}_F$ for the double-Fermi two-neutrino double-beta decay mode
as function of the ratio $4\kappa/G$  for a set of parameters (\ref{parameters}). Exact results are indicated 
with a solid line. The results obtained within the perturbation theory up to the first and second 
order in isotensor contribution to Hamiltonian are shown with dash-dotted and and dashed lines, respectively.
The restoration of isospin symmetry is achieved for $4\kappa/G=1$.
\label{fermiplot}}
\end{center}
\end{figure}

If we restrict our consideration to the first order perturbation theory, 
for transition $|4'4> \rightarrow |2'2>$ the double Fermi matrix element 
can be written as
\begin{eqnarray}
M^{2\nu}_F &\simeq& \frac{\bra{4 2}T^-\ket{4 3}\bra{4 3}T^-\ket{4 4}}
{16\chi +\frac{7}{3}\lf G+2\kappa\rh}\times\nonumber\\
&& \sqrt{\frac{2}{3}}\Omega\lf G-4\kappa 
\rh\frac{\bra{42}[A^{\dg}\tilde{A}]_0^2\ket{22}}
{\lf 28\chi+14/3\lf G+2\kappa \rh\rh}.
\end{eqnarray} 

In Fig. (\ref{fermiplot}) $M^{2\nu}_F$ is plotted as function of ratio $4\kappa/G$. We see that results 
obtained with the second order perturbation theory agree well with exact results within 
a large range of this parameter. We note also that close to a point of restoration of isospin symmetry 
($4\kappa/G=1$) a consideration of the first order perturbation theory seems to be sufficient, in 
particular for $M^{2\nu}_F \le 0.3$.

%
\section{Energy weighted sum rule of $\Delta$Z=2 nuclei}
%

We suggest that a quantity relevant for the $2\nu\beta\beta$-decay  might be the energy weighted 
double Fermi (or Gamow-Teller) sum rule associated with $\Delta$Z=2 nuclei:
\begin{eqnarray} \label{energysum}
&& S^{ew}_{F,GT}(i,f) \nonumber\\
&& = \sum_{n}(E_n - \frac{E_i + E_f}{2})\bra{f} {\cal O}_{F,GT} \ket{n}\bra{n} {\cal O}_{F,GT}\ket{i}
\nonumber\\
&& = \frac{1}{2} \bra{f}\left[ {\cal O}_{F,GT},\left[H, {\cal O}_{F,GT}\right]\right]\ket{i}.
\end{eqnarray}
Here, $|i>$ and $|f>$ are assumed to be a ground state of the initial and a ground state or
an excited state of the final nuclei participating in double-beta decay. If there is a dominance 
of contribution of a single or few states of the intermediate nucleus 
the left-hand side  of Eq. (\ref{energysum}) might be determined phenomenologically. Then,
by a calculation of the right-hand side of Eq. (\ref{energysum}) within a nuclear model  
the strengths of the residual interaction of Hamiltonian can be properly adjusted. We note that 
as the double commutator
connect states with $\Delta$Z=2 the explicit dependence on single-particle part of nuclear 
Hamiltonian is eliminated unlike it is in the case of energy weighted sum rules related 
to a single nuclear ground state. 
We note that the energy weighted double Gamow-Teller 
sum rule associated with the $2\nu\beta\beta$-decay  was discussed within the proton-neutron 
QRPA in \cite{vadim05,vadim11}.
 
We analyze the above sum rule for Fermi transitions and Hamiltonian (\ref{hamupraveny}) with $G_p=G_n$ 
within the SO(5) model. By rewriting the Hamiltonian as
\begin{eqnarray}
H&=&(e_p+e_n) N/2 + (e_p-e_n)T_z+2\chi T^-T^+\nonumber\\
&-&2G\Omega \lf A^{\dg}(-1)A(-1)+ A^{\dg}(1)A(1)\rh \nonumber\\
&-&4\kappa\Omega A^{\dg}(0)A(0)
\end{eqnarray}
and exploiting the commutation relations of the SO(5) group (\ref{comrul}) 
we find 
\begin{eqnarray} \label{dolz}
&& S^{ew}_F(i,f) = \frac{1}{2}\bra{f}\left[T^-,\left[H,T^-\right]\right]\ket{i}
\nonumber\\
&& = 2 \Omega(G-4\kappa)\bra{i}[A^{\dg}\tilde{A}]_{2}^2\ket{f} + 2\chi \bra{f}T^-T^-\ket{i}.
\nonumber\\
\end{eqnarray}

\begin{figure}[!t]
\begin{center}
\includegraphics[width=85mm]{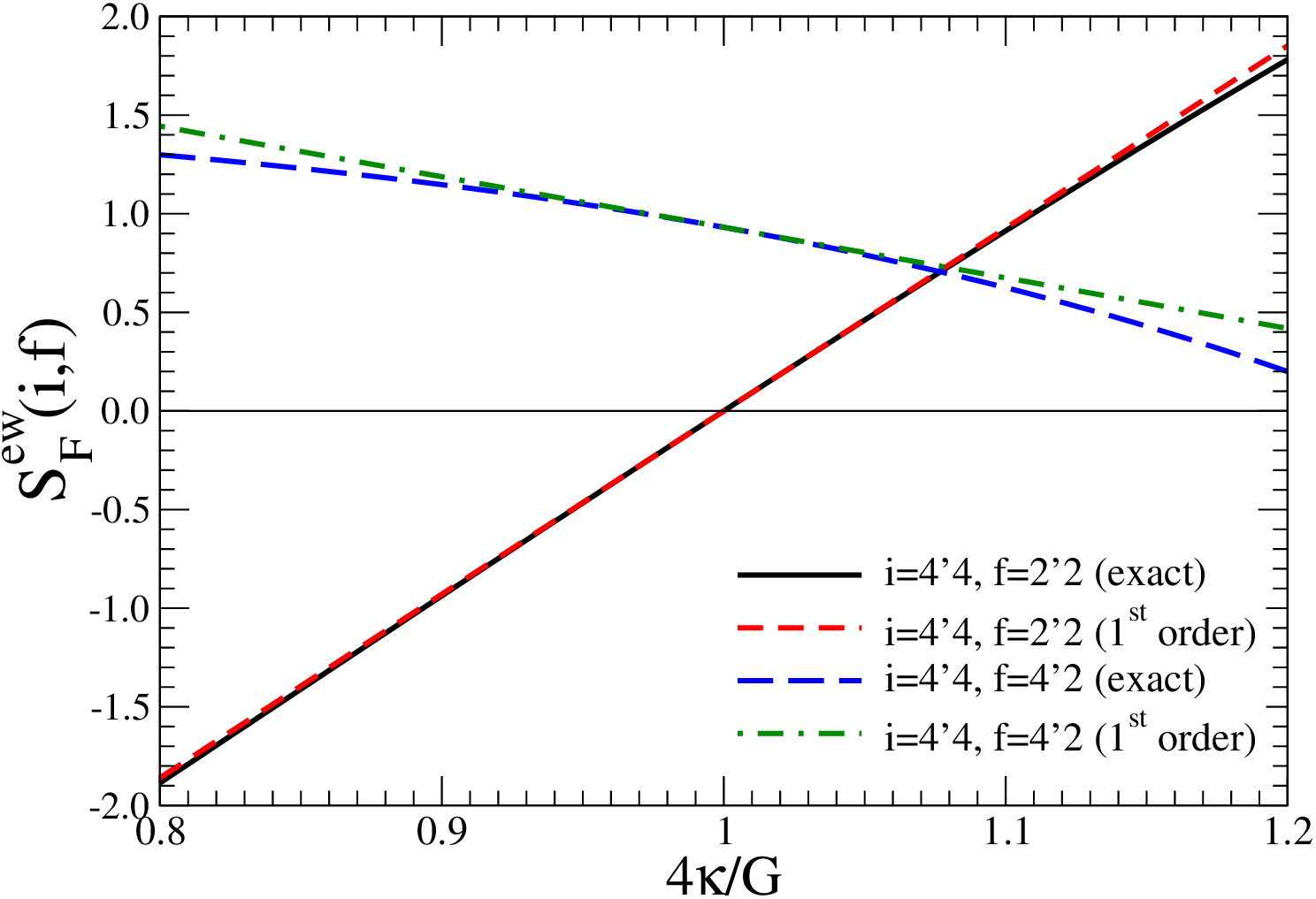}
\caption{(Color online) The energy weighted sum rule 
$S^{ew}_{F,GT}(i,f)$ (\ref{energysum}) for two sets of states 
($i=4'4$, $f=2'2$ and $i=4'4$, $f=4'2$) as function of the ratio $4\kappa/G$  
for a set of parameters (\ref{parameters}). The exact results are compared with 
those obtained within the first order perturbation theory.
\label{ewsr}}
\end{center}
\end{figure}

{\it i) The case $\ket{i}=\ket{4'4}$, $\ket{f}=\ket{2'2}$.} We have
\begin{eqnarray} 
&& S^{ew}_F(4'4,2'2)  \nonumber\\
&&= \sum_{T'}(E'_{T3} - \frac{E'_{44} + E'_{22}}{2})\bra{2'2} T^- \ket{T'3}\bra{T'3} T^-\ket{4'4}
\nonumber\\
&& = 2\Omega(G-4\kappa)\bra{4'4}[A^{\dg}\tilde{A}]_{2}^2\ket{2'2}+2\chi \bra{2'2}T^-T^-\ket{4'4}.
\label{ewsr22}\nonumber\\
\end{eqnarray}
If the first order perturbation theory is applied to any of two expressions
for energy weighted sum rule in (\ref{ewsr22}) we find
\begin{eqnarray} 
&& S^{ew}_F(4'4,2'2) \simeq  
\left( 16\chi +\frac{7}{3}\lf G+2\kappa\rh\right)\times 
\nonumber\\
&&\sqrt{\frac{2}{3}}\Omega\lf G-4\kappa \rh
\frac{\bra{42}[A^{\dg}\tilde{A}]_0^2\ket{22}}
{\lf 28\chi+14/3\lf G+2\kappa \rh\rh}\times \nonumber\\
&& \bra{42}T^{-}\ket{43}\bra{43}T^{-}\ket{44}. 
\end{eqnarray}
By comparing this expression with Eqs. (\ref{betaminus}), (\ref{betaplus})
and (\ref{enemen})  we see that only the lowest intermediate state 
$|4'3>$ contributes to the sum rule within the considered approximation.
We find again a combination of energies of involved states to be 
a  function of pairing, particle-particle and particle-hole interactions: 
$E'_{43} - (E'_{44}+E'_{22})/2 \simeq 16\chi +\frac{7}{3}\lf G+2\kappa\rh $.

{\it ii) The case $\ket{i}=\ket{4'4}$, $\ket{f}=\ket{4'2}$.} The energy
weighted sum rule is given by
\begin{eqnarray} 
&& S^{ew}_F(4'4,4'2)  \nonumber\\
&& = \sum_{T'}(E'_{T3} - \frac{E'_{44} + E'_{42}}{2})\bra{4'2} T^- \ket{T'3}\bra{T'3} T^-\ket{4'4}
\nonumber\\
&& = 2\Omega(G-4\kappa)\bra{4'4}[A^{\dg}\tilde{A}]_{2}^2\ket{4'2} 
+ 2\chi \bra{4'2}T^-T^-\ket{4'4}.
\label{ewsr42}\nonumber\\
\end{eqnarray}
Within the first order perturbation theory we find 
\begin{eqnarray} 
&& S^{ew}_F(4'4,4'2) \simeq  \left( 2\chi + \right.\nonumber\\
&& \sqrt{1/6}\Omega (4\kappa-G)
\left[ 2\bra{43}[A^{\dg}\tilde{A}]_0^2\ket{43}\right. \nonumber\\
&&\left.\left. -\bra{44}[A^{\dg}\tilde{A}]_0^2\ket{44}
-\bra{42}[A^{\dg}\tilde{A}]_0^2\ket{42}\right]\right) \nonumber\\
&& \bra{42}T^{-}\ket{43}\bra{43}T^{-}\ket{44}. 
\end{eqnarray}
We note that the dominant contribution to $S^{ew}_F(4'4,4'2)$  
comes from the transition through the single intermediate state
$|43'>$ again. For a combination of energies of involved states
we have
\begin{eqnarray}
&&E'_{43} - (E'_{44}+E'_{42})/2 =\nonumber\\
&& 2\chi+\sqrt{1/6}\Omega (4\kappa-G)
\left( 2\bra{43}[A^{\dg}\tilde{A}]_0^2\ket{43}\right. \nonumber\\
&&\left. -\bra{44}[A^{\dg}\tilde{A}]_0^2\ket{44}
-\bra{42}[A^{\dg}\tilde{A}]_0^2\ket{42}\right).
\end{eqnarray}
Thus, the energy weighted sum rule $S^{ew}_F(4'4,4'2)$ implies 
another useful relation between energies of states and 
nucleon-nucleon interactions.

In Fig. \ref{ewsr} two different energy weighted sum rules 
associated with final states $|2'2>$ and $|4'2>$ 
are plotted as function of the ratio $4\kappa/G$ for a considered 
set of parameters (\ref{parameters}). They exhibit different 
dependence on $4\kappa/G$. It is because the final 
state $|4'2>$ belongs ($|2'2>$ does not belong) 
to the same isospin multiplet as the initial nucleus.
We see a very good agreement between the exact results and 
results obtained within the first order perturbation theory,
which allows only the lowest intermediate state $|4'3>$
to contribute to a sum rule. A better agreement would be achieved
if the corresponding combination of energies of states would 
evaluated up to the second order perturbation theory. We note that
a contribution from the second lowest intermediate state to the
sum rules $S^{ew}_F(4'4,2'2)$ and $S^{ew}_F(4'4,4'2)$ appears only
in the third order perturbation theory.

%
\section{Conclusions}
%

An exactly solvable model for the description of the $2\nu\beta\beta$-decay 
processes of the Fermi type was used to discuss the
dependence of the double-beta decay matrix element $M^{2\nu}_F$ 
on different components of the residual interaction, namely
like-nucleon pairing, particle-particle and particle hole  
proton-neutron interactions. We note that the model is 
equivalent to a complete shell-model treatment in a 
single-j shell for the adopted Hamiltonian. In addition,
it reproduces the main features of the results obtained 
in realistic calculations. 

Good isospin forbids the $2\nu\beta\beta$-decay.
One needs an isotensor force to mix $\Delta$T = 2. Naturally,
the Coulomb interaction contains such a isotensor force. In 
our case we break isospin symmetry by hand. The only isospin 
violation comes from the difference of the proton-proton ($G_p$)  
and the neutron-neutron ($G_n$) pairing force compared to the 
proton-neutron isospin = 1 pairing force ($\kappa$). 
By taking the advantage of the perturbation theory up to the
second order in the isotensor contribution to the Hamiltonian 
a dominance of a contribution through a single state of the 
intermediate nucleus to $M^{2\nu}_F$ and explicite dependence 
of $M^{2\nu}_F$ on different types of nucleon-nucleon interactions 
was found. The mean-field part of  Hamiltonian
does not enter explicitly in this decomposition of double 
Fermi matrix element and is related only to the 
calculation of unperturbated states of Hamiltonian.  

Further, the importance of the energy weighted sum rule 
associated with $\Delta$Z=2 nuclei for fitting 
different components of residual interaction
of the Hamiltonian was pointed out. It goes without saying 
that a further studies, in particular by considering 
realistic nuclear Hamiltonian and Gamow-Teller transitions, 
are of great interest.

\acknowledgments

This work is supported in part by the Deutsche Forschungsgemeinschaft
within the project "Nuclear matrix elements of Neutrino Physics and Cosmology"
FA67/40-1 and  by the grant of the Ministry of Education
and Science of the Russian Federation (contract 12.741.12.0150).
F. \v S. acknowledges the support by the VEGA Grant agency
of the Slovak Republic under the contract No. 1/0876/12 and by the Ministry 
of Education, Youth and Sports of the Czech Republic under contract LM2011027.

%
\appendix
\section{The SO(5) algebra and matrix elements \label{ap}}
%

Following \cite{parikh} we introduce operators of the SO(5) group,
which are expressed with operators (\ref{operators}) as follows:
\begin{eqnarray}
\begin{array}{cc}
H_1=N/2-\Omega, &H_2=T_Z,  \\
E_{11}=\sqrt{\Omega}A\dg(1),&E_{-1-1}=\sqrt{\Omega}A(1), \\
E_{1-1}=-\sqrt{\Omega}A\dg(-1),& E_{-11}=-\sqrt{\Omega}A(-1), \\
E_{10}=\sqrt{\Omega}A\dg(0),& E_{-10}=\sqrt{\Omega}A(0), \\
E_{01}=\frac{1}{2}\sqrt{2}T^{+},& E_{0-1}=\frac{1}{2}\sqrt{2}T^-, \\
 \end{array} \nonumber
\end{eqnarray}
Their commutation relations are \cite{parikh}
\begin{eqnarray}\label{comrul}
&&\left[H_1,H_2\right]=0, \quad \left[H_1,E_{a b}\right]
=a E_{a b},\nonumber \\ 
&&  \left[H_2,E_{a b}\right]=b E_{a b}, \nonumber \\
&& [E_{a b},E_{-a -b}]=a H_1+ b H_2 \nonumber
\end{eqnarray}
and
\begin{eqnarray}
&& [E_{a b},E_{a^{\prime} b^{\prime}}]=  
\pm E_{a + a^{\prime} b + b^{\prime} }, 
\end{eqnarray}
if $a + a^{\prime} = 0,\pm1$ and $b + b^{\prime} = 0,\pm1$. Otherwise,
$[E_{a b},E_{a^{\prime} b^{\prime}}]= 0$. 

For the present task, states with  with seniority s=0 are considered.
Thus, it is sufficient to define them with quantum numbers 
N, T and T$_z$. They are constructed with help of the isospin 
lowering operator T$^-$ on the state $|{\rm N, T, T_z=T}\rangle$,
which is given by \cite{parikh}
\begin{eqnarray}
\ket{NTT} &=& N(a,b) O^a_+ O^b_{00} \ket{N=4\Omega, T=T_z=0}, \nonumber
\end{eqnarray}
with
\begin{eqnarray}
O_+    &=& E_{-1 1}, \nonumber\\
O_{00} &=&  2 E_{-1 1} E_{-1 -1} + E_{-1 0} E_{-1 0}
\end{eqnarray}
$O_+$ reduces the number of particles by two units and increases
the isospin by one unit and $O_{00}$ reduces the number of particles by four units.
 $a$ and $b$ are integers:
\begin{eqnarray}
 a = T, ~~~~~~ b = \Omega - \frac{T}{2} - \frac{N}{4}.
\end{eqnarray}
From a construction of the states it follows that a difference in 
isospin of two states with fixed $N,T_z$ is an even number.

The reduced matrix elements are calculated with help of the 
Wigner-Eckart theorem in the convention as follows:
\be
\left\langle T^{\prime} T_z^{\prime}\left|T^p_{q}\right|T T_z \right\rangle
= C^{T^{\prime} T_z^{\prime}}_{TT_zpq}\left\langle T^{\prime} 
\left|\left|T^p\right|\right|T \right\rangle
\end{equation}

Particular Clebsh-Gordan coefficients of interest are given by \cite{va}
\begin{eqnarray}
C_{T T_z 20}^{T T_z}=
\frac{3T_z^2-T(T+1)}{\sqrt{(2T-1)T(T+1)(2T+3)}} \nonumber
\end{eqnarray}
\begin{widetext}
\begin{eqnarray}
C_{T T_z 20}^{T+2 T_z}=
\sqrt{\frac{3(T+T_z+1)(T+T_z+2)(T-T_z+1)(T-T_z+2)}{(2T+1)(2T+2)(2T+3)(T+2)}} \nonumber
\end{eqnarray}

We present relevant reduced matrix elements, which agree with those of \cite{hirsch97} 
up to few corrections:
\be
\left\langle T+2 \left|\left|[A^{\dg}\tilde{A}]^2\right|\right|T  \right\rangle
= - \frac{1}{2\Omega}\sqrt{\frac{\left(T+2\right)\left(T+N/2+3\right)
\left(2\Omega-T-N/2\right)\left(T+1\right)\left(N/2-T\right)
\left(2\Omega+T-N/2+3\right)}{\left(2T+3\right)\left(2T+5\right)}} \nonumber
\end{equation}
\begin{eqnarray}
\left\langle T \left|\left|[A^{\dg}\tilde{A}]^2\right|\right|T\right\rangle
&=&\frac{1}{\sqrt{6}C^{TT}_{TT20}}
\left[\left\langle N T T\left|A^{\dg}(1)A(1)\right|N T T \right\rangle + 
\left\langle N T T\left|A^{\dg}(-1)A(-1)\right|N T T \right\rangle
\right. \nonumber \\
&-&\left. 2\left\langle N T T\left|A^{\dg}(0)A(0)\right|N T T \right\rangle\right] \nonumber
\end{eqnarray}

\begin{eqnarray}
\left\langle T \left|\left|[A^{\dg}\tilde{A}]^1\right|\right|T\right\rangle
=\frac{1}{\sqrt{2}C^{TT}_{TT10}}
\left[\left\langle N T T\left|A^{\dg}(-1)A(-1)\right|N T T \right\rangle
-\left\langle N T T\left|A^{\dg}(1)A(1)\right|N T T \right\rangle \right] \nonumber
\end{eqnarray}

\begin{eqnarray}
\left\langle N T T\left|A^{\dg}(1)A(1)\right|N T T \right\rangle= 
\frac{1}{\Omega}\left[-\Omega+T+N/2+\frac{\left(2\Omega-T-N/2\right)
\left(T+N/2+3\right)\left(T+1\right)}{2\left(2T+3\right)}\right]\label{jj}\nonumber
\end{eqnarray}
\be\left\langle N T T\left|A^{\dg}(-1)A(-1)\right|N T T \right\rangle 
= \frac{1}{\Omega}\left[\frac{\left(2\Omega+T-N/2+3\right)\left(-T+N/2\right)
\left(T+1\right)}{2\left(2T+3\right)}\right]\nonumber
\end{equation}
\begin{eqnarray} \left\langle N T T\left|A^{\dg}(0)A(0)\right|N T T \right\rangle= 
&&\frac{1}{\Omega}\left[-\Omega+N/2+\frac{\left(2\Omega-T-N/2\right)\left(T+N/2+3\right)\Omega}
{\left(2\Omega+T-N/2+1\right)\left(-T+N/2+2\right)}\times \right.\nonumber \\ 
&&\left. \left\langle N+4 T T\left|A^{\dg}(0)A(0)\right|N+4 T T \right\rangle \right] 
\label{lastnme}
\end{eqnarray}
\end{widetext}
The matrix element on the right hand side of Eq. (\ref{lastnme})
can be calculated recurrently by keeping in mind that 
for $N_{max}=4\Omega -2T$ we have 
\be
\left\langle N_{max} T T\left|A^{\dg}(0)A(0)\right|N_{max} T T
\right\rangle=1-T/\Omega \label{nmax}\nonumber 
\end{equation} 
For isospin raising (lowering) operators the Condon Shortley 
convention is assumed:
\be
T^{\pm}\ket{N,T,T_z}=\sqrt{\lf T\pm T_z+1 \rh \lf T\mp T_z
\rh}\ket{N,T,T_z\pm1}. \nonumber 
\end{equation}

%

%


\begin{thebibliography}{99}

%
%

\bibitem{doi83} M. Doi, T. Kotani, and E. Tagasugi, Prog. Theor. Phys. (Supp.) {\bf 83}, 1 (1985).

\bibitem{haxton} W.C. Haxton and G.S. Stephenson Jr., Prog. Part. Nucl. Phys. {\bf 12}, 409 (1984).

\bibitem{suci98} J. Suhonen and O. Civitarese, Phys. Rep. {\bf 300}, 123 (1998).

\bibitem{fs98} A. Faessler and F. {\v S}imkovic, J. Phys. G {\bf 24}, 2139  (1998).

\bibitem{ves12} J.D. Vergados, H. Ejiri, and F. \v Simkovic, Rep. Prog. Phys. {\bf 75}, 106301 (2012).

\bibitem{recbb} A.S. Barabash, Phys. Rev. C {\bf 81}, 035501  (2010).

\bibitem{domin} P. Domin, S. Kovalenko, F. {\v S}imkovic, and S.V. Semenov, Nucl. Phys. A {\bf 753}, 337 (2005).

\bibitem{rath07} S. Singh, R. Chandra, P.K. Rath, P.K. Raina, and J.G. Hirsch, Eur. Phys. J. A {\bf 33}, 375 (2007).

\bibitem{sarr} R. Alvarez-Rodriguez, P. Sarriguren, E. Moya de Guerra, L. Pacearescu, A. Faessler, and F. \v Simkovic,
       Phys. Rev. C {\bf 70}, 064309 (2004). 

\bibitem{poves12} E. Caurier, F. Nowacki, and A. Poves, Phys. Lett. B {\bf 711}, 62 (2012).

\bibitem{vogel86} P. Vogel and M.R. Zirnbauer, Phys. Rev. Lett. {\bf 57}, 3148 (1986).

\bibitem{civitarese87} O. Civitarese, A. Faessler, and T. Tomoda, Phys. Lett. B {\bf  149}, 11 (1987).

\bibitem{muto89} K. Muto, E. Bender, and H.V. Klapdor, Z. Phys. A {\bf 334}, 177 (1989). 

\bibitem{newpar13} F. \v Simkovic, V. Rodin, A. Faessler, and P. Vogel, Phys. Rev. C  {\bf 87}, 045501 (2013).

\bibitem{desplan90} J. Bernabeu, B. Desplanques, J. Navarro, and S. Noguera, Z. Phys. C {\bf 46}, 323 (1990).

\bibitem{kuzmin} V.A. Kuz'min and V.G. Soloviev,  Nucl. Phys. A {\bf 486}, 118 (1988).

\bibitem{muto92} K. Muto, E. Bender, T. Oda, and H. V. Klapdor-Kleingrothaus, Z. Phys. A {\bf 341}, 407 (1992).

\bibitem{civi94} O. Civitarese and J. Suhonen, J. Phys. G {\bf 20}, 1441 (1994).

\bibitem{hplb} J. G. Hirsch, P.O. Hess, and O. Civitarese, Phys. Lett. B {\bf 390}, 36 (1997).

\bibitem{hirsch97} J.G. Hirsch, P.O. Hess, and O. Civitarese, Phys. Rev. C {\bf 56}, 199 (1997).

\bibitem{krmpotic98} F. Krmpoti\'{c}, E.J.V. de Passos, D.S. Delion, J. Dukelsky, and P. Schuck,
    Nucl. Phys. A {\bf 637}, 295 (1998). 

\bibitem{engel04} J. Engel and P. Vogel, Phys. Rev. C {\bf 69}, 034304 (2004).

\bibitem{anatomy} F. \v Simkovic, A. Faessler, P. Vogel, J. Engel, 
   Phys. Rev. C {\bf 77}, 045503 (2008).

\bibitem{engel12} J. Engel, J. Phys. G {\bf 39}, 124001 (2012).

\bibitem{parikh} J.C. Parikh, Nucl. Phys. {\bf 63}, 214 (1965).

\bibitem{vadim05} V.A. Rodin, M.H. Urin, and A. Faessler, Nucl. Phys. A  {\bf 748}, 295 (2005).

\bibitem{vadim11} V. Rodin and A. Faessler, Phys. Rev. C {\bf 84}, 014322 (2011). 

\bibitem{va} D.A.~Varshalovich, A.N.~Moskalev, and
V.K.~Khersonskii, \textit{Kvantovaja teorija uglovovo momenta}, (Izdatelstvo Nauka, Leningrad, 1975).

\end{thebibliography}
\end{document}